# Effect of white LED DC-bias on modulation speed for visible light communications

**Peng Deng*** **and Mohsen Kavehrad**

*Department of Electrical Engineering, The Pennsylvania State University, University Park, PA 16802, USA*
*pxd18@psu.edu

**Abstract:** The light emitting diode (LED) nonlinearities distortion induced degradation in the performance of visible light communication (VLC) systems can be controlled by optimizing the DC bias point of the LED. In this paper, we theoretically analyze and experimentally demonstrate the effect of white LED DC bias on nonlinear modulation bandwidth and dynamic range of the VLC system. The linear dynamic range is enhanced by using series-connected LED chips, and the modulation bandwidth is extended to 40 MHz by post-equalization without using a blue filter. The experimental results well match the theoretical model of LED nonlinear modulation characteristics. The results show that the modulation bandwidth increases and saturates with an increase in LED DC bias current due to nonlinear effect of carrier lifetime and junction capacitance. The optimized DC-bias current that corresponds to the minimum BER increases with the increase of data rate. A 60-Mbps NRZ transmission can be achieved with BER threshold of $10^{-3}$ by properly adjusting LED DC bias point.

## 1. Introduction

Visible light communications (VLC) based on light-emitting diodes (LEDs) merges lighting and data communications in applications due to their energy efficiency, small size and reliability[1, 2]. Visible light spectrum provides a potential solution to the looming radio frequency (RF) spectrum crisis for high speed wireless communications[1]. VLC using lighting LEDs has attracted more and more attention since it can offer many advantages over RF such as THz license-free bandwidth, free of electromagnetic interference (EMI), and integration into lighting infrastructure.

However, the phosphor-based white-light LED has a limited modulation bandwidth of about a few megahertz that is caused by the slow response time of phosphor. It restricts the direct modulation speed of VLC systems[3]. There are several approaches to improve the modulation bandwidth, such as using a blue filter to filter out the slow yellow component[3], and equalization of the driving circuitry[4]. Using an optical filter will increase the complexity and reduce the received optical power. Despite the bandwidth limitation of the LEDs, techniques such as high-order modulation/multiplexing[5, 6], optical multi-input multi-output (MIMO) [7-9] and parallel data transmission have enabled communications of up to 1 Gb/s using a single white light phosphor LED[3, 10] and up to 3.4 Gb/s using RGB LEDs[11]. These approaches have low bit-error-rate (BER) and high complexity in comparison with simple on-off-key (OOK) direct modulation.

Moreover, VLC systems also suffer from the nonlinear modulation characteristics and the limited linear dynamic range of LEDs that are determined by the DC bias points of LEDs[12]. For an ideal LED, if the injected carriers arrive instantaneously at the diffusion region, the rise time of the spontaneous emission is governed solely by the differential lifetime of the carriers[13]. However, for driving an LED, the junction capacitance cause delay in the arrival time of the injected carriers at the recombination region[14]. Thus, the modulation bandwidth and rise time of the spontaneous emission in an LED will be either material-limited by the differential lifetime $\tau_s$ or circuit-limited by the time constant $\tau_c$ due to the junction capacitance of LED[15]. In fact, LEDs are non-linear devices whose parasitic resistance and space-charge capacitance depend strongly on the DC bias current. These induce nonlinear electrical-optical properties and nonlinear modulation characteristics of LEDs that depend on the DC bias[16]. The limited linear dynamic range of LEDs will cause clipping noise and signal distortion. The limited modulation bandwidth will cause inter symbol interference noise and data rate degradation. The LED nonlinear modulation characteristics lead to distortion in signal-noise-ratio (SNR) degradation in bit-error-rate (BER) performance of VLC system[17]. Therefore, it's necessary to investigate the effect of LED DC bias on nonlinear modulation characteristics and optimize the DC bias to control the nonlinear effect and improve the VLC system performance.

In this paper, we theoretically analyze and experimentally demonstrate the effect of white LED DC bias on nonlinear modulation characteristics of the VLC system, including nonlinear modulation bandwidth and nonlinear electrical-optical properties related to DC bias current. The experimental results well match the theoretical model of LED nonlinear modulation characteristics. We use the series-connected UWLED chips to enhance linear dynamic range, and the RC post-equalization without a blue filter to extend the modulation bandwidth to 40 MHz. The results show that VLC system bandwidth increases with LED bias current, and a 60-Mbps NRZ transmission is achievable with a BER threshold of $10^{-3}$. Furthermore, the optimized DC-bias current that corresponds to minimum BER increases with an increase in data rate. This can help in controlling the dimmable illumination brightness of white LEDs to match the modulation speed and enhance system performance.

## 2. Nonlinear modulation characteristics of white LED

### *2.1 LED modulation bandwidth due to carrier lifetime*

The differential carrier lifetimes in InGaN-based LED can be described by the simple ABC model. This rate equation model considers that the current through the device is made up of three contributions: a non-radiative current $I_A$ due to Shockley-Read-Hall recombination at defect sites, a current $I_B$ due to radiative recombination of electrons and holes, and an Auger current $I_C$ due to cubic non-radiative recombination. The total current is thus[13]

$$I = I_A + I_B + I_C = qsd(AN + BN^2 + CN^3) \tag{1}$$

where $I$ is the operation current, $q$ is the elementary charge, $s$ and $d$ are the device area and total thickness of the quantum wells (QWs), $N$ is carrier density, $A$ , $B$ and $C$ are the coefficients of Shockley-Read-Hall (SRH), radiative recombination and Auger non-radiative recombination. This simple model is made under the assumption that $A$ , $B$ and $C$ are independent of $N$, and the carrier density is the same in all QWs. In fact, the radiative coefficient $B$ decreases with the carrier density $N$ due to a reduction in the optical matrix element for the inter-band transitions. An internal quantum efficiency is thus[18]

$$IQE = \frac{BN^2}{AN + BN^2 + CN^3} \tag{2}$$

Under small-signal modulation conditions, the luminescence decay is mono-exponential with the differential time constant. The differential lifetime is then given by the derivative of the recombination rate with respect to carrier density[18]

$$\frac{1}{\tau_s} = A + BN + CN^2 \tag{3}$$

In the case that linear and quadratic terms dominate the recombination rate, the simplified expression between carrier life time and bias current can be derived

$$\frac{1}{\tau_s^2} = A^2 + \frac{4B}{qV} I \tag{4}$$

Thus, the LED electrical-optical 3-dB bandwidth related to the differential lifetime can be obtained

$$f_s = \frac{1}{2\pi\tau_s} = \frac{1}{2\pi}\sqrt{A^2 + \frac{4B}{qV} I} \tag{5}$$

*2.2 LED modulation bandwidth due to space-charge capacitance*

The common feature of LED is that the large junction over the entire wafer presents a space-charge capacitance of the junction in parallel with the diffusion capacitance of carrier lifetime. Figure 1 shows the equivalent circuit for the transient behavior of LED. In addition to the diffusion capacitance $C_d$ of carrier lifetime, there is the space-charge capacitance $Cs$ of the depletion layer. The space-charge capacitance is given by[14]

$$C_s = \frac{C_0}{(1 - v_d / \phi)^m}$$

where $C_0 = sc_0$ , $c_0$ is the zero-bias capacitance per unit area, $\varnothing$ is the barrier voltage, the exponent $m$ is 1/2. The LED resistance $R_s$ in the equivalent circuit accounts for the series resistance of the bulk material and the contact resistances. Thus, the time constant due to the space-charge capacitance is given by

$$\tau_c = C_s R_s = \frac{R_s C_0}{(1 - v_d / \phi)^m} \tag{6}$$

The electrical-optical 3-dB bandwidth related to space-charge capacitance can be obtained

$$f_c = \frac{1}{2\pi\tau_c} = \frac{(1 - v_d / \phi)^m}{2\pi R_s C_0} \tag{7}$$

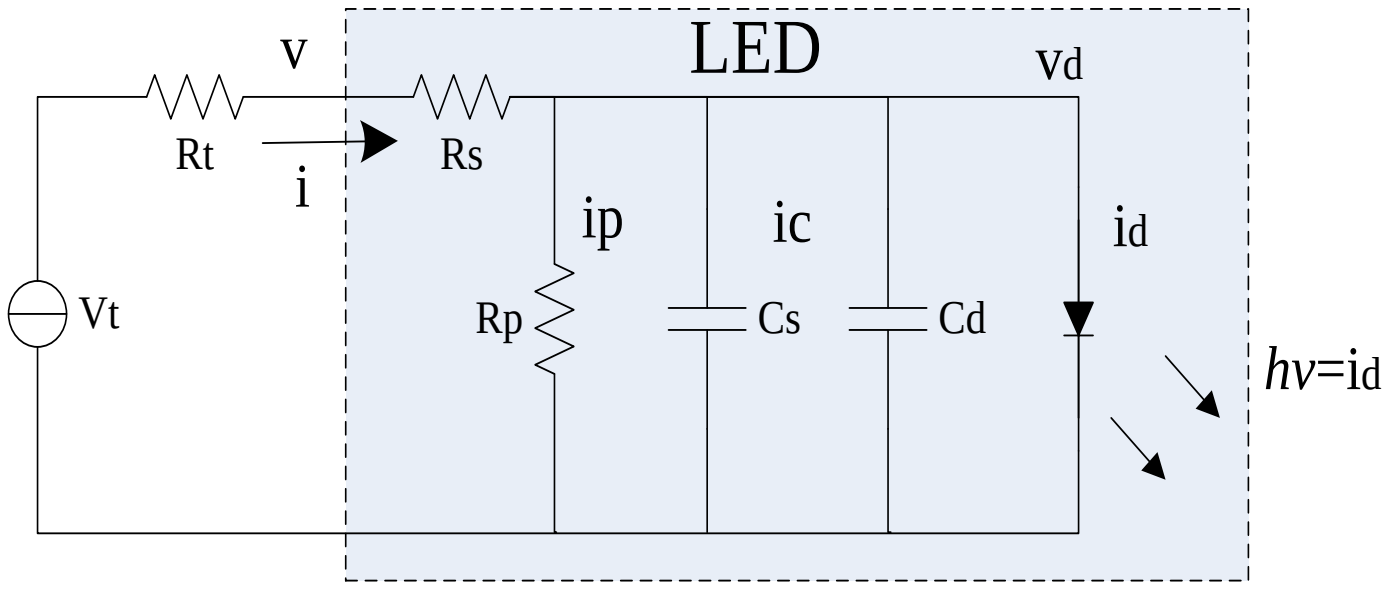

Fig. 1. The equivalent circuit for the transient behavior of LED.

*2.3 Nonlinear Electrical-Optical characteristics of LED*

LEDs are non-linear devices such as the series resistance and capacitance depend strongly on the bias current as shown in Fig. 1. A nonlinear resistor represents the *I-V* characteristic of the LED by the Shockley equation[14]

$$i_d = I_0 \left[ \exp\left( \frac{qv_d}{nkT} \right) - 1 \right] \tag{8}$$

where $k$ is the Boltzmann's constant, $q$ is the electron charge, T is junction temperature, $n$ is diode ideality factor, $v_d$ is the junction voltage. The factor $n$ is approximately 2 for an InGaN LED. $I_0$ is the reverse saturation current defined by[16]

$$I_0 = E_0 \exp\left( -E_g / kT \right) \tag{9}$$

where the band-gap energy $E_g = qv_{th}$, $v_{th}$ is the threshold voltage, $E_0$ is a parameter related to the junction material size. The emission optical intensity $P$ as a function of energy $E$ is given by

$$P(E) \propto \sqrt{E - E_g}\, \exp\left[-E / (kT)\right] \tag{10}$$

The LED nonlinear *I–V* characteristic, as given by the Shockley equation, needs to be modified in order to take into account parasitic resistances. A series resistance can be caused by excessive contact resistance or by the resistance of the neutral regions. A parallel resistance can be caused by any channel that bypasses the p-n junction. This bypass can be caused by damaged regions of the p-n junction or by surface imperfections. Assuming a shunt with resistance $R_P$ (parallel to the ideal diode) and a series resistance $R_S$ (in series with the ideal diode and the shunt), the *I–V* characteristic of a forward-biased p-n junction diode($v_d = v - IR_s$, $v_d >> kT/q$) is given by[16]

$$I - \frac{(v - IR_s)}{R_p} = I_0 \exp\left[q(v - IR_s)/(nkT)\right] \tag{11}$$

For devices with a high parallel resistance ($R_P \to \infty$), the diode *I–V* characteristic can be written as

$$I = I_0 \exp\left[e(v - IR_s)/(nkT)\right] \tag{12}$$

For $R_P \to \infty$ and $R_S \to 0$, this equation reduces to the Shockley equation.

### *2.4 Equivalent LED modulation bandwidth related to DC bias current*

Considering the nonlinear electrical-optical characteristics of LED, insert Eq.(12) into Eq.(6), the total LED rise time due to carrier spontaneous life time and space-charge capacitance can be defined by

$$\tau_{LED}(I) = \tau_s + \tau_c = \frac{1}{\sqrt{A^2 + 4BI/qV}} + \frac{R_s C_0}{\sqrt{1 - \ln(I/I_0)nkT/q\phi}} \tag{13}$$

The total equivalent electrical-optical 3-dB bandwidth due to the differential lifetime and space-charge capacitance can be expressed by the LED bias current

$$f_{LED}(I) = \frac{1}{2\pi(\tau_s + \tau_c)} = \frac{1}{2\pi} \frac{\sqrt{(A^2 + 4BI/qV)[1 - \ln(I/I_0)nkT/q\phi]}}{\sqrt{1 - \ln(I/I_0)nkT/q\phi} + R_s C_0 \sqrt{A^2 + 4BI/qV}} \tag{14}$$

## 3. System Description

### *3.1 Experiment setup*

Figure 2 presents a diagram of the experimental VLC system set-up. At the analogue transmitter front-end, the output signal from the network analyzer (HP 4396B) or function generator (AWG 4438c ESG) was first amplified with the purpose to increase the LED modulation depth and then superimposed onto the LED DC bias current by the aid of a bias-tee (HP 11612A). The DC current biased the white LED in its linear region and also provided sufficient average optical power for illumination. The light source was a phosphorescent white LED (Osram OSTAR), devised for general lighting. This module consists of four chips, each providing a luminous flux of 520 lm with a 120° full opening angle when driving at 700 mA. An aspheric lens was fixed to make sure the light transmits along the regular direction. The modulated light emitted from the white LED was transmitted over an indoor VLC channel.

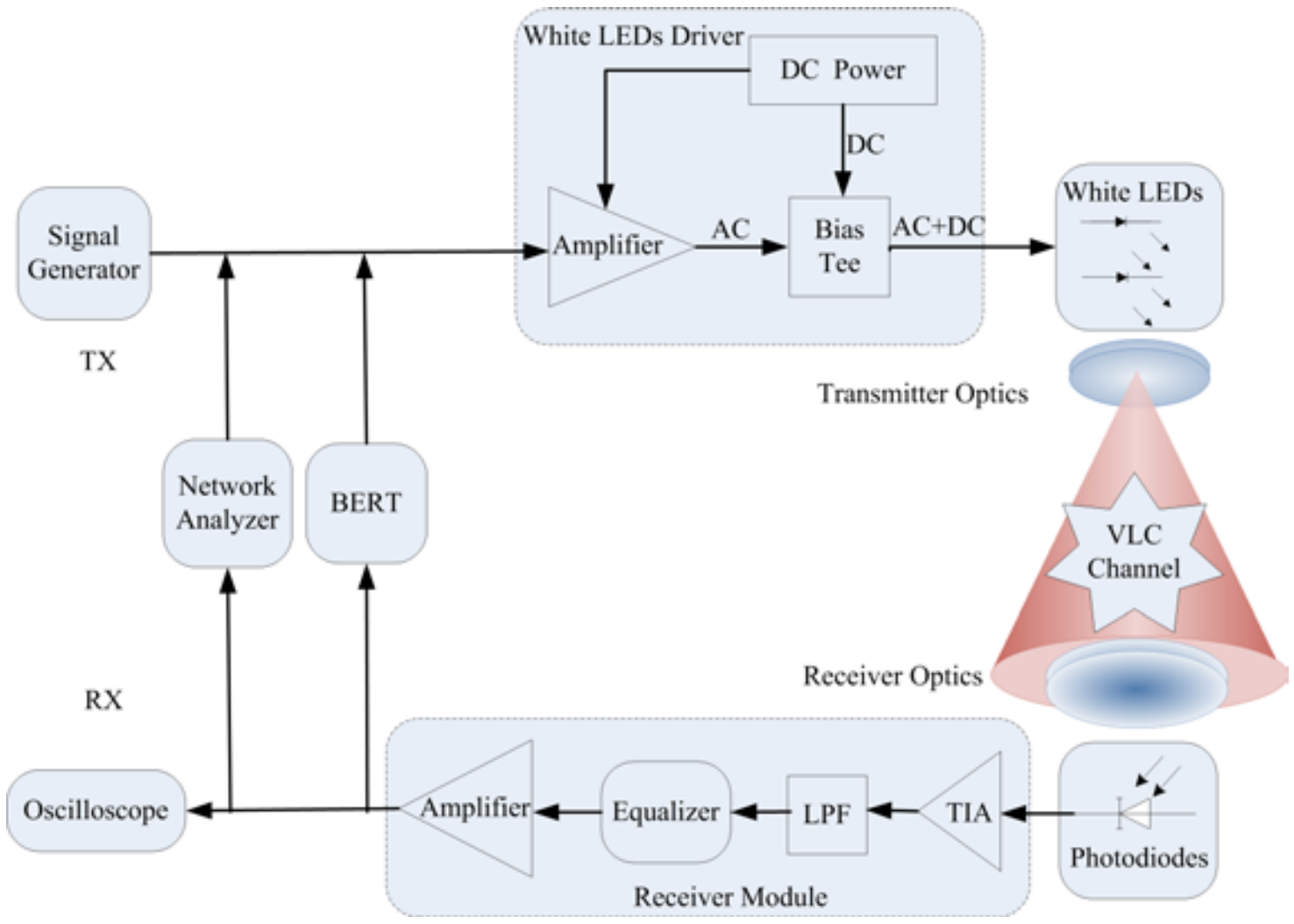

Fig. 2. Experimental setup of Visible Light Communication system using White LEDs.

Light from the white LED was imaged onto a high-speed PIN photodetector (PDA10A) through an aspheric convex lens (25 cm focal length). The photodetector has a detection wavelength range of 2000-1100 nm with a responsivity of 0.45 A/W and an active area of 0.8 $mm^2$. It has a bandwidth of 150 MHz and a root mean square (RMS) noise of 1.5 mV. The optical signal from the LED was converted to electrical current signal through the PIN photodiode and then the current signal was amplified to voltage signal by a low noise transimpedance amplifier (TIA). The low-pass filter with a cutoff frequency of 150 MHz can reduce the high-frequency noise. A first-order analogue post-equalization was developed to extend the VLC system bandwidth based on phosphorescent white LED. The received electrical signal was then amplified by a wideband coaxial amplifier, which was connected to a digital oscilloscope(HP 4396B) and a bit error rate tester (BERT).

### *3.2 LED DC-bias and dynamic range*

We measured optical spectrum of different phosphor white LED chips (Osram Ostar) as show in Fig. 3. It can be seen that a commercial warm white LED (CWLED) contains only 10% blue component of the overall emitted power, while the emitted ultra white LED light (UWLED) consists of 41% blue component and less slow yellow component from the phosphor. As the modulation bandwidth of blue light is higher than that of yellow component, the ultra white LED (UWLED) can be used to improve the modulation efficiency and bandwidth for high speed visible light communications.

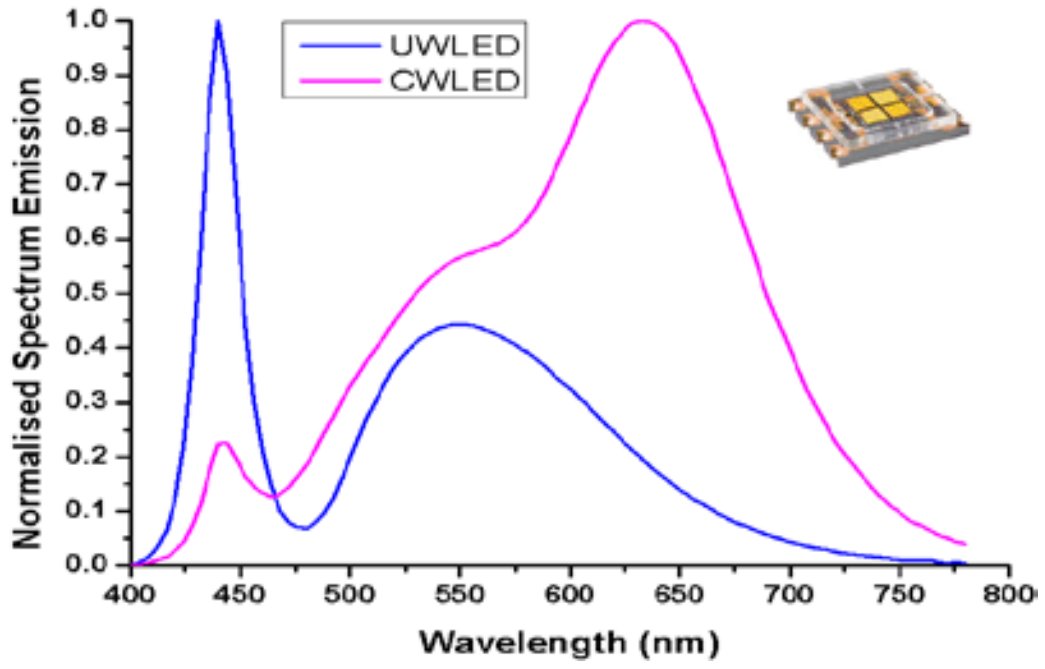

Fig. 3. The measured optical spectrum of different White LED chips, inset: Osram White LED.

We measured the electrical-optical properties of several LED chips. Figure 4(a) shows the experimental and theoretical DC-bias current $I$ against forward voltage $v$ for a single LED.

Considering the nonlinear series resistances $R_s$ and parallel resistance $R_p$ of LED in Eq.(11), the measured LED bias current $I$ against the junction voltage $v_d$ characteristic well match theoretical nonlinear $I$-$v_d$ model. Figure 4(b) shows output optical power against forward voltage for a single LED. The LED output light is cut off below the threshold 10mA @ 2.7V. Above the threshold voltage, the LED can output a linear optical power around the optimal bias point of 450mA @ 3.3V within the modulation dynamic range of 600mV. However, the electrical-to-optical conversion efficiency saturated and decreased, this because the internal quantum efficiency of LED reduced and nonlinear thermal effect appeared when bias current and forward voltage beyond 1100mA @ 3.7V. The DC and AC currents must be optimized to ensure that the LED chip does not cutoff or overheat that leads to degradation and nonlinearity in output optical power.

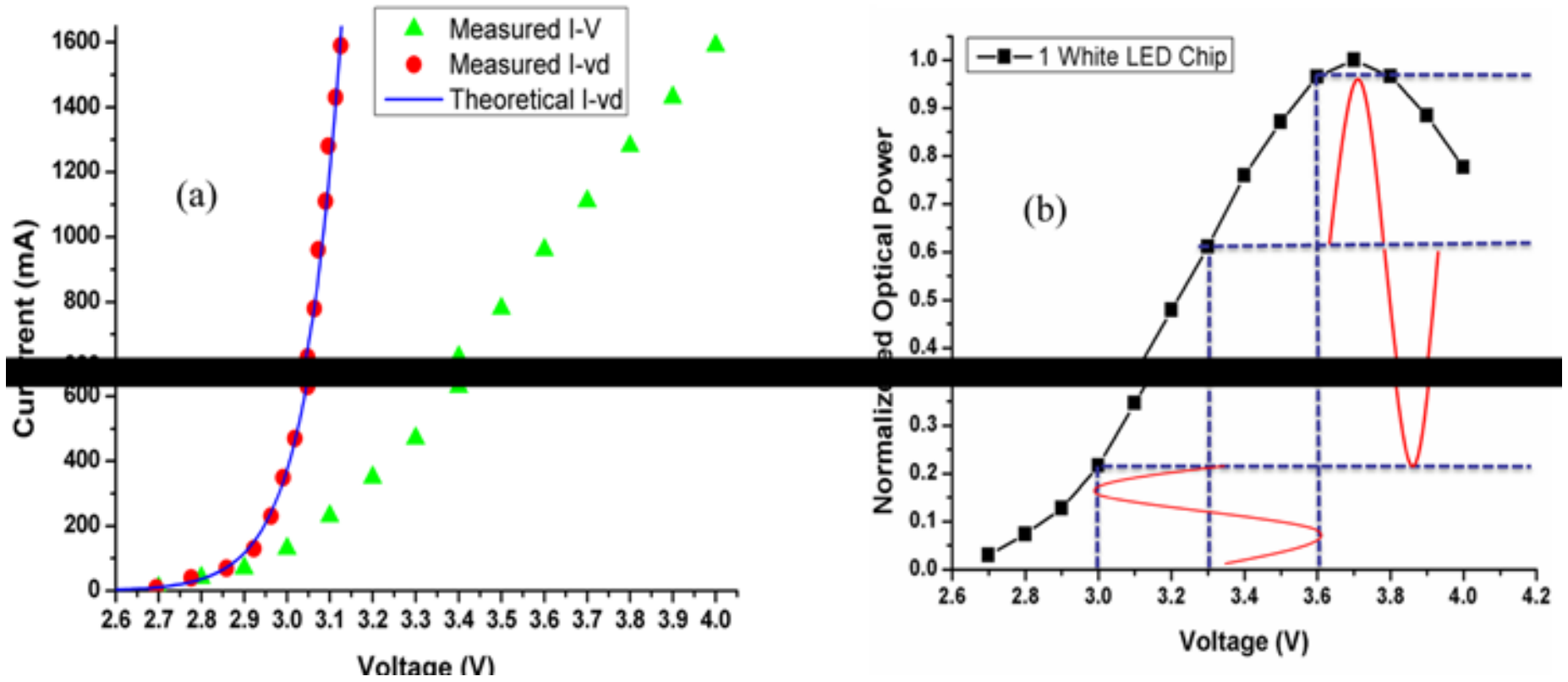

Fig. 4. The electrical to optical properties of different white LED chips. (a) DC-bias current against forward voltage for single LED. (b) Normalized optical power against forward voltage for single LED.

The DC-bias current and output optical power against forward voltage for two parallel-connected and series-connected LED chips are shown in Fig. 5. It can be seen that nonlinear region of parallel-connected LED chips fluctuates dramatically above 3000 mA @ 3.5 V due to the larger forward current thermal effect, whereas the linear modulation dynamic range of the series-connected LED chips increases to 1200 mV around the optimal DC-bias point 370 mA @ 6.4V due to the improved impedance. Thus, the series-connected LED chips can enhance the linear dynamic range.

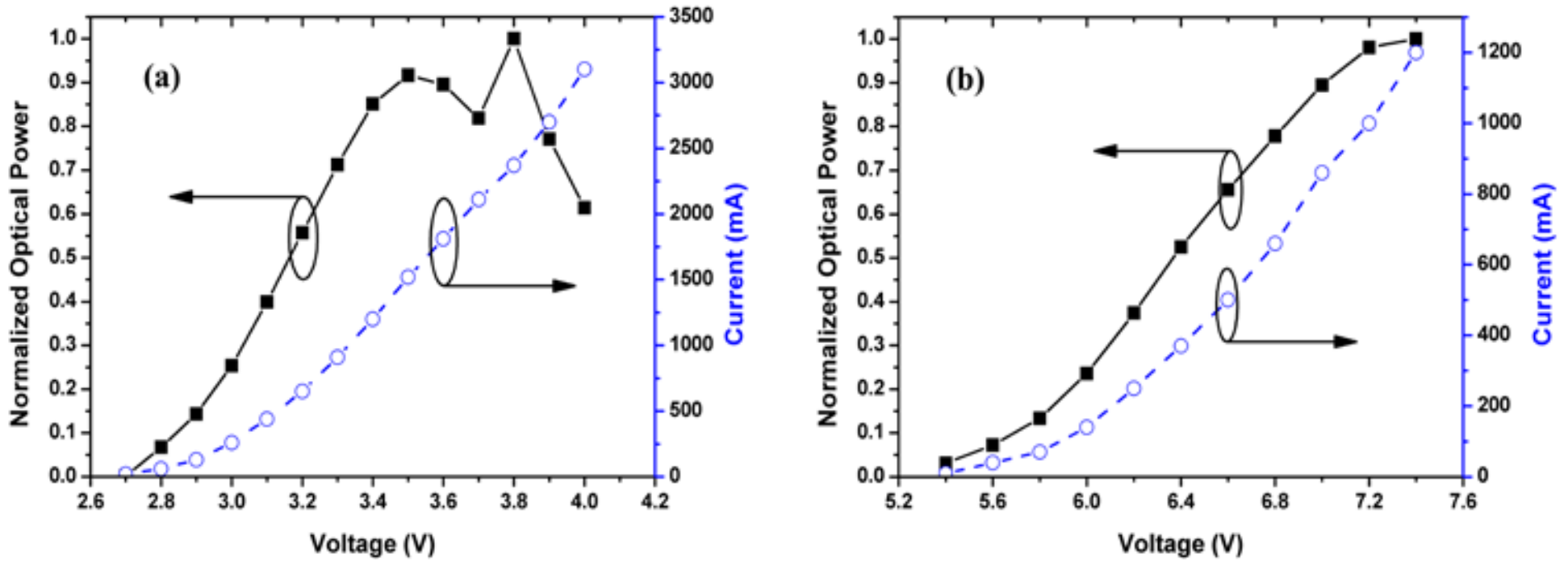

Fig. 5. (a) DC-bias current and normalized optical power against forward voltage for two parallel connected LEDs. (b) DC-bias current and normalized optical power against forward voltage for two series connected LEDs.

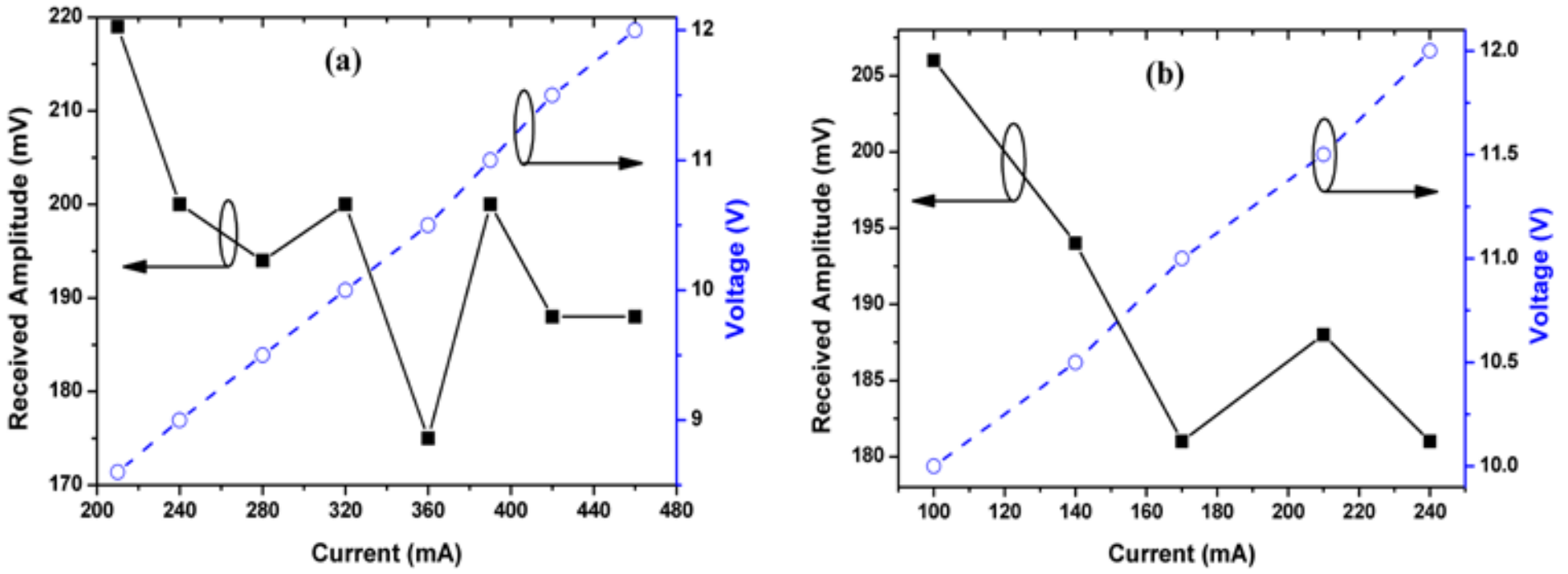

Fig. 6. The DC-bias current and voltage effect on VLC system response for (a) two series-connected LED chips, (b) three series-connected LED Chips. Transmitter signal amplitude is 100 mV.

We evaluated VLC transceiver system response at the frequency of 10 MHz along the distance of 1 meter. Using the same transmitter signal amplitude of 100 mV, we measured the receive signal amplitude for different LED DC-bias power levels. The effect of DC forward voltage and current on system AC signal response for two and three series-connected LED chips are shown in Fig. 6. It is shown that the received signal and system response fluctuate to decline gradually with the increase of LED DC-bias current and voltage, because high DC power will lead to the nonlinear thermal effect that saturated output optical power and degraded AC signal amplitude. As compared, larger number of series-connected LED chips can mitigate the nonlinear fluctuation of system response.

## 4. Results and Discussion

We first evaluate the effect of LED DC bias power on frequency response and system bandwidth, and then demonstrate DC power effect on VLC data transmission. The configuration of the VLC system demonstration is shown in Fig. 2. A directed line of sight (LOS) VLC link operates over a distance of 1 m resulting from the deployment of a single LED chip, but this can be extended by using an array of LED chips.

### *4.1 Effect of LED DC-bias on system bandwidth*

The effect of LED DC-bias current on electrical-optical-electrical (EOE) channel frequency response of VLC transceiver is shown in Fig. 7. The magnitude of the channel frequency response was measured by a network analyzer. It can be seen that the magnitude of frequency response reduced when bias current is less than 100 mA due to clipping of the lower peaks at the turn on point. Furthermore, VLC system frequency response magnitude decreased with an increase in LED DC bias current for frequencies smaller than 10 MHz, because the high magnitude of low frequency at the large bias current will lead to saturation and nonlinear thermal effect on electrical-to-optical conversion efficiency. However, for high frequency signals in Fig. 7(b), the system response improves with the increasing DC bias current due to higher optical power detected at the receiver.

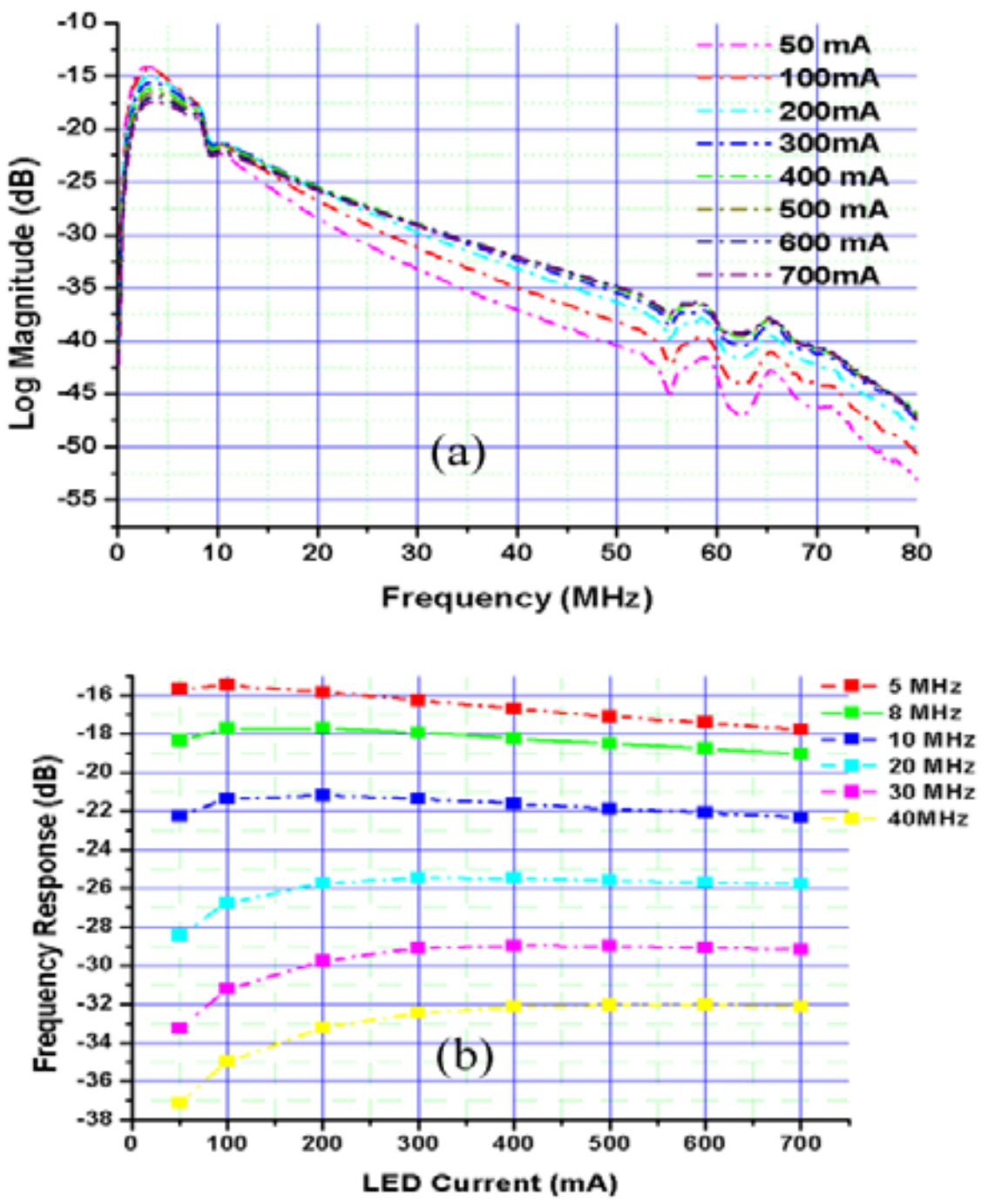

Fig. 7. (a) The measured EOE channel frequency response of VLC transceiver for different LED DC bias currents. (b) Frequency response versus LED DC bias current for different frequency.

The experimental and theoretical relationship between 3-dB bandwidth of the VLC system and LED bias current are compared in Fig. 8. It is shown that the 3-dB bandwidth of the VLC system increases with LED bias current, because the bandwidth due to carrier lifetime *fs* in Eq. (5) increases significantly with carrier density and driving current. Furthermore, the LED bandwidth saturates and approaches to 7 MHz when DC bias current is larger than 250 mA, because the bandwidth due to space-charge capacitance $f_c$ in Eq. (7) reduces apparently with bias current. It is shown that experimental measurement of LED bias current related equivalent bandwidth well match the theoretical results in Eq.(14). Since the receiver itself has a relatively large bandwidth (photo-detector 150 MHz, amplifier 1 GHz), the system bandwidth (7 MHz) is limited by the LED module.

A first-order equalizer that consists of a capacitor in parallel with a resistor (R=1 kΩ, C= 30 pF) is developed to enhance the system bandwidth. Figure 9 presents the frequency response of the equalizer and the corresponding equalized VLC transceiver system. It can be seen that the optimized analogue equalizer can extend the bandwidth of the VLC transceiver from 7 MHz to 40 MHz.

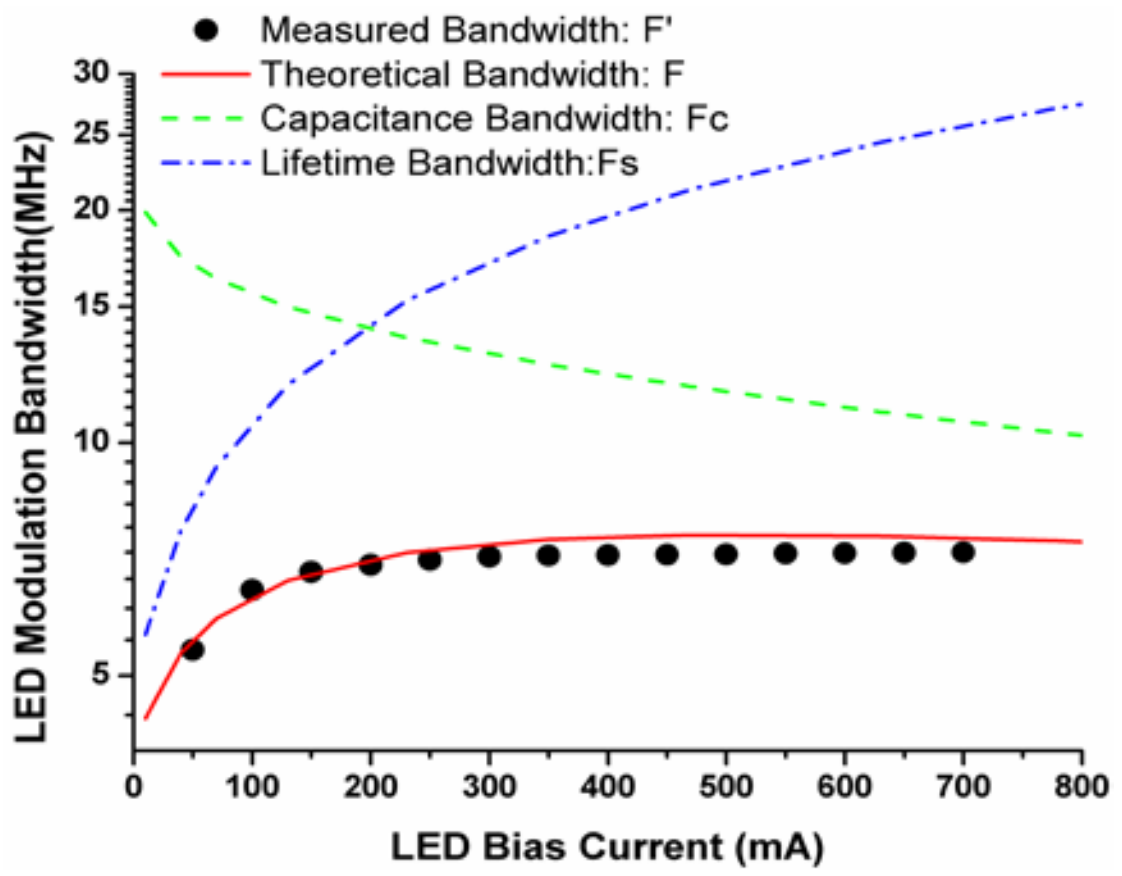

Fig. 8. The relationship between 3-dB bandwidth and average LED current.

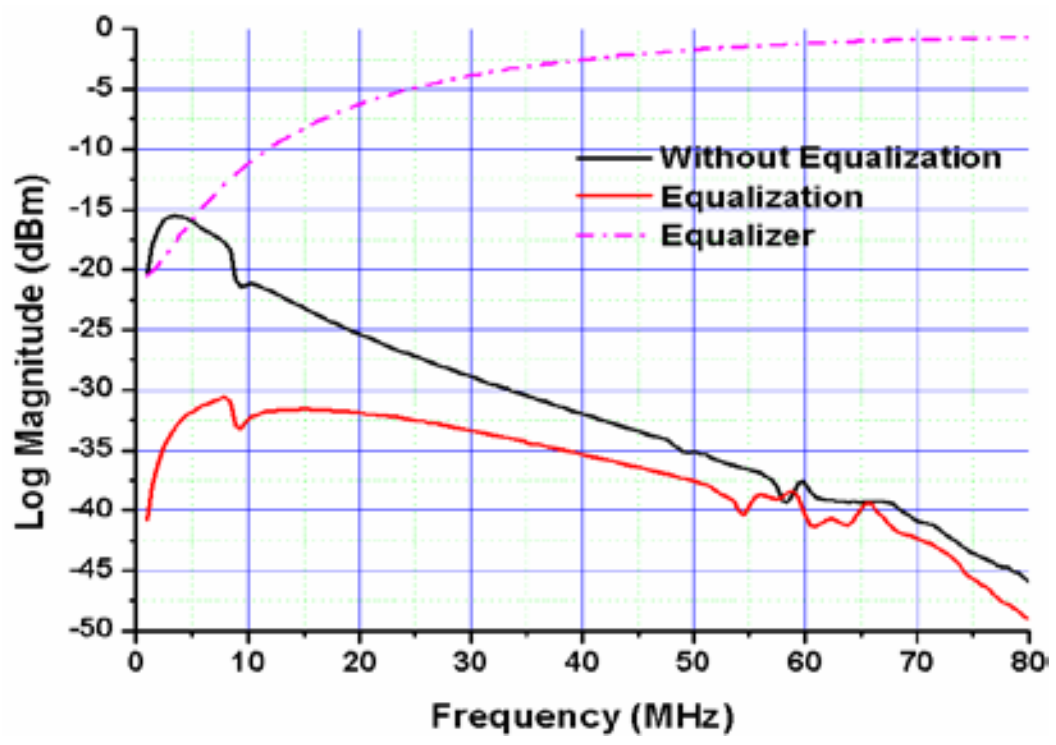

Fig. 9. Electrical-Optical-Electrical system frequency response of VLC transceiver without equalization, VLC transceiver with equalization and the analogue equalizer.

### *4.2 Effect of LED DC-bias on data Rate and BER*

A pseudorandom binary sequence (PRBS)-10 OOK-NRZ data stream with a peak-to-peak voltage swing of 2.5 V is used to modulate a single LED light. Figure 10 shows the relationship between the measured bit-error-rate(BER) performance and date rate of the equalized VLC system for different DC bias currents. It can be seen that the maximum date rate at DC bias current of 510 mA can achieve 60 Mbps for a BER threshold of $10^{-3}$. In contrast, BER performance degraded obviously for both low DC bias current of 200 mA and high DC bias current of 680 mA.

The effect of LED DC bias current on VLC BER performance for different data rates is illustrated in Fig. 11. It can be seen that the maximum date rate at the DC bias current of 510 mA can achieve 60 Mbps for a BER threshold of $10^{-3}$. For low transmission data rates (20 Mbps, 32 Mbps), the high BER observed at the small bias current of less than 100 mA. This is mainly due to clipping of the lower peak at the turn on threshold and the relatively low output optical power. The DC bias current that corresponds to minimal BER value is used to identify the optimum bias point, where the LED is operating in a linear region with less probability of clipping the signal peaks. The high BER observed at the large bias current more than 200 mA is mainly due to nonlinear thermal effect and the clipping of the upper peak at the maximum output optical power. For high transmission data rates (40 Mbps, 60 Mbps), BER decreases to a minimum value with the increase of LED bias current. This is because relatively high levels of optical power are required to overcome the received signal-to-noise penalty at high date rate. Therefore, the optimized DC-bias current of an LED driver that corresponds to minimum BER increases gradually with an increase in data rate.

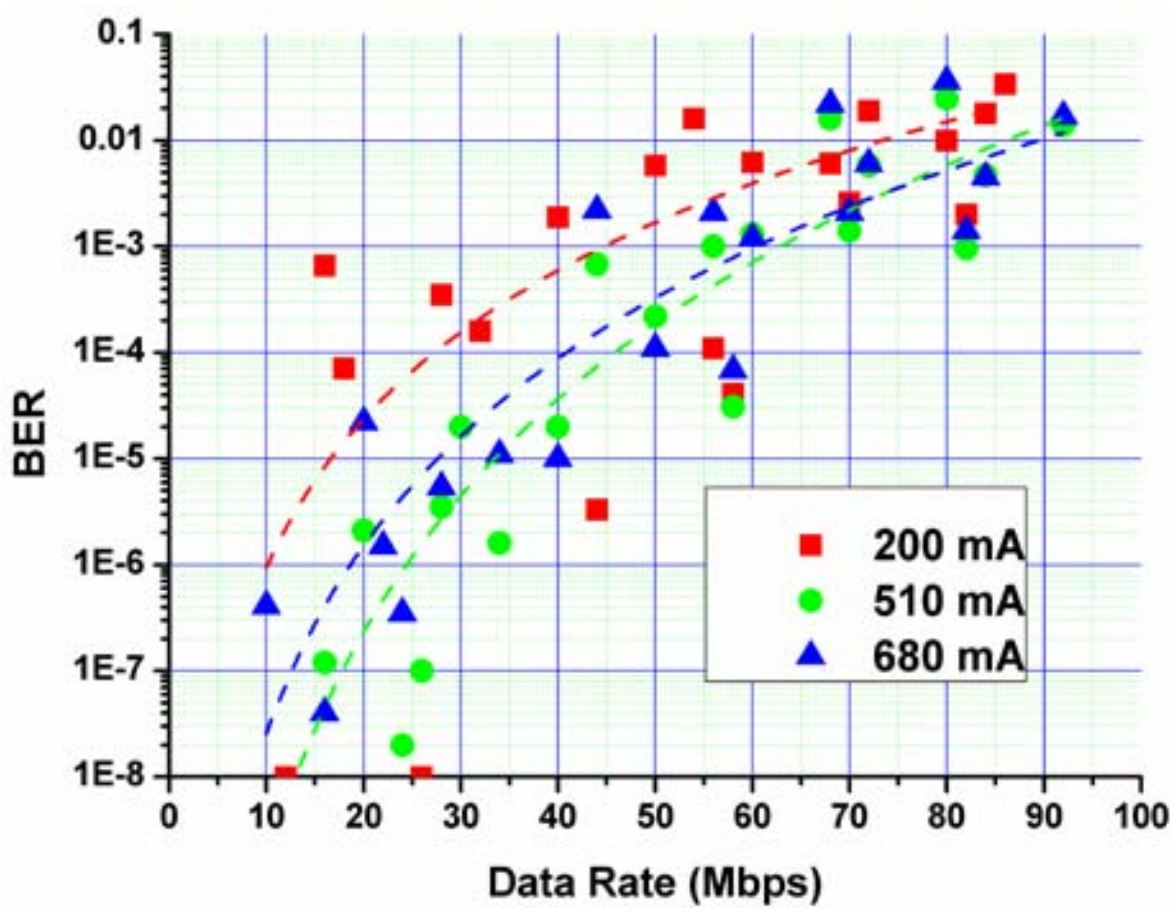

Fig. 10. The relationship between VLC bit error rate and date rate for different LED DC-bias currents.

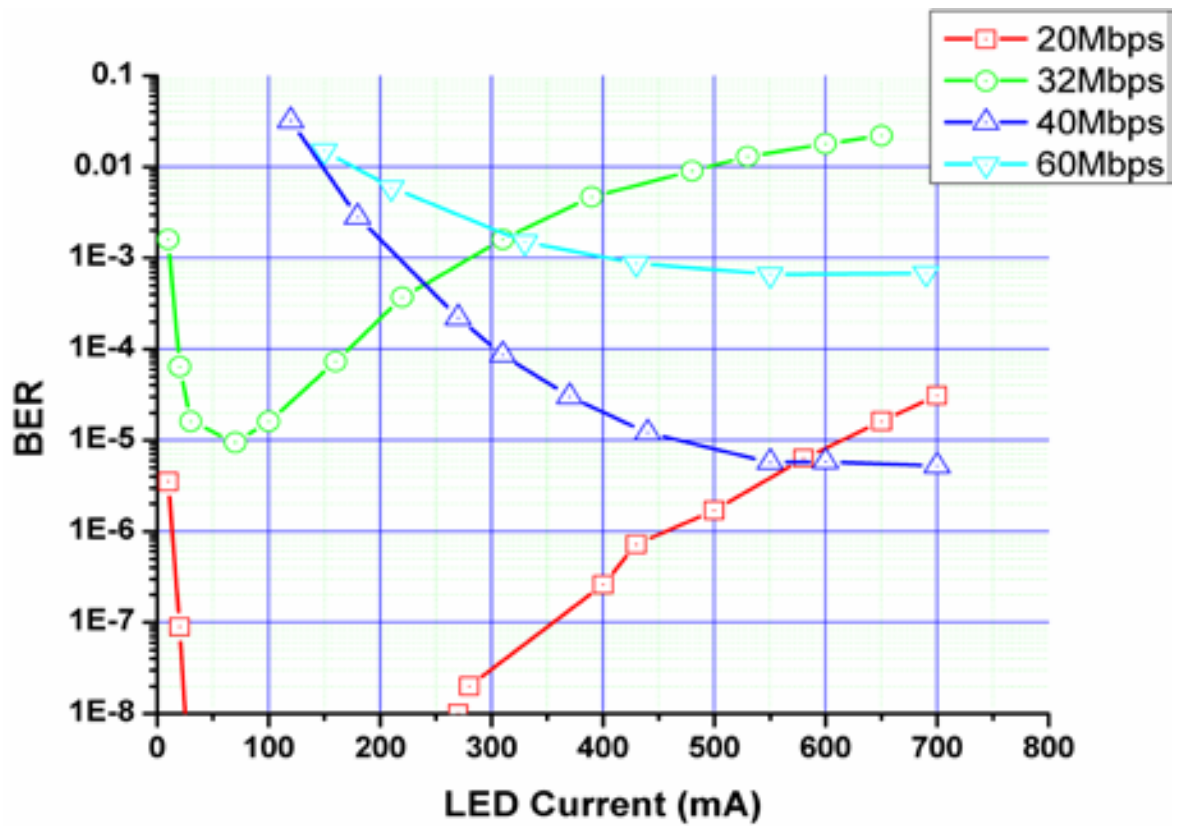

Fig. 11. The relationship between VLC bit error rate and LED current for different date rate.

## 5. Conclusions

In this paper, we have theoretically analyzed and experimentally demonstrated the effect of LED DC bias on modulation bandwidth and data rate of visible light communications system, considering the LED nonlinear modulation characteristics due to carrier lifetime and junction capacitance. We used the series-connected ultra white LED chips to enhance linear dynamic range, and the RC post-equalization without a blue filter to extend the modulation bandwidth to 40 MHz. The experimental results well match the theoretical model of LED nonlinear modulation characteristics. The measurement results show that VLC system bandwidth increases and saturates with an increase in DC bias current due to carrier lifetime and junction capacitance of LED. Data transmission in VLC system shows that a 60-Mbps NRZ transmission can be achieved with a BER threshold of $10^{-3}$ by properly adjusting the LED DC bias point. Furthermore, the optimized DC-bias current of the LED driver that corresponds to minimum BER increases with the increase of data rate. The modulation bandwidth can be enhanced by reducing parasitic resistances and capacitances of LED, or carrier sweep-out of the active region. This can help in controlling the dimmable illumination brightness of white LEDs to match the modulation speed and improve system performance.

## Acknowledgments

This research was financially supported by the Pennsylvania State University CICTR COWA (http://cowa.psu.edu), through US National Science foundation (NSF) ECCS Directorate under award (1201636).